# On The Hydrosphere Stability of *TESS* Targets: Applications to 700 d, 256 b, and 203 b

Paul Bonney[1,2], Julia Kennefick[2]

ABSTRACT

A main determinant of the habitability of exoplanets is the presence of stable liquid surface water. In an era of abundant possible targets, the potential to find a habitable world remains a driving force in prioritization. We present here a data-forward method to investigate the likelihood of a stable hydrosphere on the timescales of the formation of life, ~1 Gyr, and beyond. As our primary application, we use this method to examine the potential hydrospheres of *TESS* Objects of Interest 700 d, 256 b, and 203 b. We first present our selection criteria, which is based on an implementation of the Earth Similarity Index, as well as the results of an initial investigation into the desiccation of the targets, which reveals that TOI 203 b is almost certainly desiccated based on *TESS* observations. We then describe the characterization of the remaining targets and their host stars from 2MASS, Gaia, and *TESS* data and the derivation of sampled probability distributions for their parameters. Following this, we describe our process of simulating the desiccation of the targets' hydrospheres using the Virtual Planet Simulator, VPLanet, with inputs directly linked to the previously derived probability distributions. We find that 50.86% of the likely cases for TOI 700 d are desiccated and no modeled cases for TOI 256 b are without water. In addition, we calculate the remaining water inventory for the targets, the percentage of cases that are continuing to lose water, and the rate at which these cases are losing water.

[1] Corresponding author pmbonney@email.uark.edu
[2] Department of Physics, University of Arkansas, 825 W. Dickson St. Fayetteville, AR 72701, USA

## 1. INTRODUCTION

The launch of the Transiting Exoplanet Survey Satelite (*TESS*) has built upon the legacy of the Kepler missions, together discovering an astonishing number of exoplanets. This, along with the recently successful launch of the James Webb Space Telescope (*JWST*), has made the potential discovery of a habitable planet outside of our solar system a more tangible goal.

Unfortunately, the current best chance of robustly assessing the habitability (i.e. stable liquid surface water) of a planet lies with *JWST*, which has significant limitations in both time available as well as resolution for smaller, potentially more Earth-like planets. As such, it is beneficial to prioritize future observations to those which are most likely to produce the most results using as few resources as possible.

Most current methods of prioritization lack proper treatment of the different potential states of the planetary systems they examine (i.e. orbital configuration and planetary composition). Instead, many study only one state or a few (e.g. Barnes et al. 2016; O'Malley-James & Kaltenegger 2017; Becker et al. 2020) often ignoring the effects of many potentially important parameters, especially if they are poorly constrained. A tantalizing option is a grid-based approach to fully capture the intricacies of the parameter space. However, this approach is slow and unwieldy, can easily ignore a number of likely states for the sake of speed, and moreover is devoid of likelihood information. Indeed, even the study by Fleming et al. (2020) uses pre-prescribed prior distributions instead of observational data.

To this end, we have created a new end-to-end data-forward method of prioritization that is fast and flexible. We utilize Monte Carlo fitting of *TESS* lightcurves to fully analyze the effect of the potential distributions of parameters such as density and insolation on modeling the planets' orbital and atmospheric evolution. This allows us to robustly and thoroughly determine which have the highest chance of being habitable at the present day.

The method leaves room for additional constraints both theoretical and observational. However we present a basic application using only *TESS* data and previously observed stellar parameters.

In this paper, we describe the selection process of our two targets, *TESS* Objects of Interest (TOIs) 256 b and 700 d as well as the initial characterization of the targets and their host stars in Section 2. Following this, Section 3 details the specifications for our coupled atmospheric and orbital evolution simulations using the modeling software package VPlanet (Barnes et al. 2020) as well as how these simulations are statistically linked to the *TESS* observations. In addition, we present four metrics to assess the degradation of a target planet's hydrosphere within the method. In Section 4, the results of the method are shown for both the initial lightcurve characterization as well as the outcome of the simulations detailed in the previous section. Finally, Section 5 presents our conclusions and recommendations for future observations of both targets.

## 2. INITIAL PLANET SELECTION & CHARACTERIZATION

Inferring meaningful statistics about the stability of liquid surface water and an Earth-like atmosphere requires accurate and complete parameter distributions for the planets' physical and orbital characteristics. In order to properly simulate these probability distributions, we begin from *TESS*'s 2-minute cadence photometry data on the targets and perform an in-depth data reduction and fit the data using a Monte Carlo method. The resultant traces of each parameter are used in the desiccation simulations. This link between the data and simulations is detailed in Section 3. The reduction of each Target Pixel File (TPF) is handled separately for each sector observation. The resultant lightcurves are then merged and fit following the procedure in section 2.3.

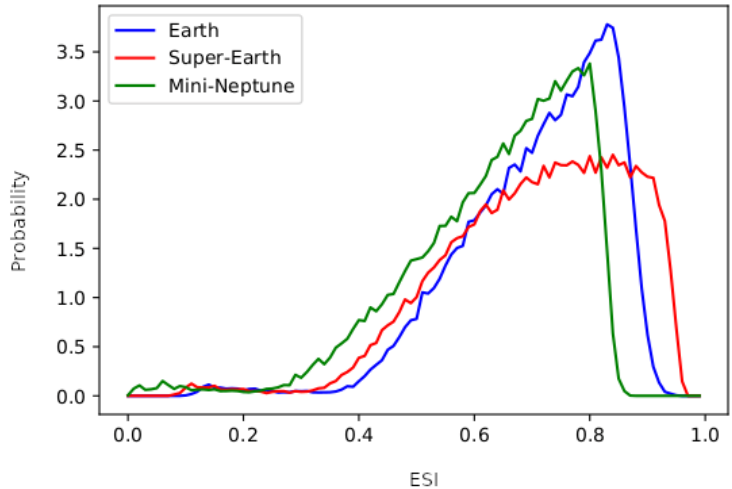

**Figure 1.** The probability of ESI values for TOI 256. The similarity to Earth (blue curve) is presented along with the similarity to a super-Earth (red curve) and a mini-Neptune (green curve) with Earth-like insolation.

*2.1: Target Selection*

We implemented an adaptation of the Earth Similarity Index (ESI; Schulze-Makuch et al. 2011) to select targets of interest. The ESI serves as a good basis for our purposes in this study by preferring Earth-like planets in terms of radius and insolation.

Instead of relying on one value for the ESI, however, we used the Markov Chain Monte Carlo (MCMC) sampler *emcee* (Foreman-Mackey et al. 2013) to estimate the likelihood that a target has an ESI of above 0.8. Input parameter distributions were obtained from the Data Validation files produced by the TESS Science Operations Pipeline (SPOC; Jenkins et al. 2016) for all TESS Objects of Interest (TOIs). The MCMC was sampled using 30 chains of 5000 steps each, resulting in 150,000 samples per parameter (Figure 1).

The ESI of each TOI was calculated using the standard formulation except for using insolation in place of surface temperature. The inner threshold of insolation was set at the values determined by Leconte et al. (2015) to cause a runaway greenhouse effect with the outer threshold being the $CO_2$ maximum greenhouse needed to sustain liquid water (Kopparapu et al. 2014). The three TOIs presented in this study were the top targets for this selection criterion. Additional TOIs have since been discovered, but none have had a higher likelihood of being Earthlike.

Initial *VPlanet* simulations were performed on each planet by sampling the system parameters available from the Exoplanet Follow-up Observing Program (Exo-FOP 2019), assuming a Gaussian distribution, and performing the method described in Section III for the primary atmosphere test. These initial simulations indicated that TOI 203b was almost certainly desiccated: 99.08% of the models that started with 10 Earth Oceans of water ended with no water. The simulations of TOI 700d and 256b, however, only returned 30.35% and 15.18% desiccated models respectively. From these results, we decided to move forward in applying the process detailed herein to TOIs 700d and 256b only.

*2.2: Host star parameterizations*

The detection and characterization of transiting exoplanets begins with the host star, and inaccurate measurements of stellar properties have effects at every level of observation and modeling. In order to ensure our targets are accurately modeled, we determine the host stars' radius, mean density, and effective temperature using standard methods in the analysis of M-dwarfs (Berta-Thompson et al. 2015; Dittmann et al. 2017; Ment at al. 2019; Kostov et al. 2019).

We acquire the measured K-band magnitude and parallax from the *2MASS* (Skrutskie et al. 2006) and *Gaia* (Gaia Collaboration et al. 2016; Gaia Collaboration et al. 2018; Luri et al., 2018) surveys, respectively. The input parameters are assumed to be Gaussian, and the uncertainties on output parameters are estimated with a simple Monte Carlo simulation using 200000 samples from the input distributions. The necessary stellar parameters as well as their estimated uncertainties are included in Table 1.

The mass of the stellar host is determined using the mass-luminosity relation from Benedict et al. (2016). This mass is then used to estimate the stellar radius using a relationship for single stars (Boyajian et al. 2012) and the mean density is calculated. We estimate the bolometric luminosity as the mean of the corrected K and V magnitudes (Mann et al. 2015, erratum and Pecaut & Mamajek 2013 respectively) and calculate the effective temperature using the Stephan-Boltzmann law. Finally, we confirm these parameters using the Mann et al. (2019) K-mag-distance-mass relationship.

*2.3: TESS data acquisition and reduction*

As with other studies using *TESS* data (e.g. Gilbert et al. 2020), we decided to apply a reduction scheme that differs from the standard pipeline. These changes consist of creating a custom photometry aperture for the files, applying reduction and decorrelation techniques while masking transits, and moving the Gaussian process fitting/removal to the data fitting portion. This process necessitates that we begin with the 2-minute cadence TPFs which have undergone an initial cleaning for cosmic rays and flagged observations.

We used the Python program *lightkurve* (Lightkurve Collaboration et al. 2018) to download the publicly available TPFs and created a custom aperture for each using the threshold method. This method selects pixels that have a flux higher than 3 standard deviations above the median brightness and are contiguous with the center pixel. Each aperture is visually inspected to ensure that it includes only the target star and is devoid of other contamination.

To avoid overcorrecting the data during transit, each TPF is then masked for the duration of each planet's transit(s) with room for error. Expected transit times and lengths as well as the variance for each are taken from Exo-FOP and are included in Table 1.

Using the pixel and time series masks, we perform a pixel-level decorrelation on each TPF using the built-in method in *lightkurve*. Finally, we extract each lightcurve from the TPFs, remove NANs and outliers $> 7\,\sigma$ outside of expected transits, and normalize the lightcurves. An example of the final, cleaned lightcurve is given in Figure 2.

Lightcurves for each sector's observation are cleaned separately. The entire catalogue of observations of TOI 256 were used, but only the first 11 sectors' observations were used for TOI 700 to be consistent with the data used in Gilbert et al. (2020) and Suissa et al. (2020).

Finally, the lightcurves for each sectors' observation are stitched together based on the average of the error in the lightcurve. The lightcurves for TOI 256 had similar error and are stitched together before fitting. As TOI 700 has some differences in error between sectors, the sectors were grouped together into 4 groups: sectors 1, 3, 4, and 5 form group 1, sectors 6 and 7 form group 2, sectors 8 and 9 form group 3, and sectors 10, 11, and 13 form group 4.

Table 1. Model Inputs

| Parameter | Value | Source |
|---|---|---|
| *TOI 256* | | |
| **Host Star** | | |
| $K_S$ (mag) | $8.821 \pm 0.023$ | 2MASS |
| $V_J$ (mag) | $13.10 \pm 0.01$ | Gilbert et al. (2020) |
| Parallax (mas) | $66.829 \pm 0.048$ | Gaia DR2 |
| $R_*$ ($R_\odot$) | $0.2097 \pm 0.00427$ | This Work |
| $M_*$ ($M_\odot$) | $0.1794 \pm 0.00244$ | This Work |
| $T_{eff}$ (K) | $3200 \pm 44$ | This Work |
| **TOI 256 b** | | |
| $T_0(BJD - 2457000)$ | $1399.93 \pm 1$ | Exo-FOP |
| $\ln(Period\ [days])$ | $3.208 \pm 1$ | Exo-FOP |
| $\ln(Transit\ Depth\ [ppm])$ | $8.503 \pm 0.055$ | Exo-FOP |
| **TOI 256 c** | | |
| $T_0(BJD - 2457000)$ | $1389.29 \pm 1$ | Exo-FOP |
| $\ln(Period\ [days])$ | $1.329 \pm 1$ | Exo-FOP |
| $\ln(Transit\ Depth\ [ppm])$ | $7.832 \pm 0.063$ | Exo-FOP |
| *TOI 700* | | |
| **Host Star** | | |
| $K_S$ (mag) | $8.634 \pm 0.022$ | 2MASS |
| $V_J$ (mag) | $14.18 \pm 0.03$ | Dittmann et al. (2017) |
| Parallax (mas) | $32.133 \pm 0.027$ | Gaia DR2 |
| $R_*$ ($R_\odot$) | $0.4403 \pm 0.00937$ | This Work |
| $M_*$ ($M_\odot$) | $0.4634 \pm 0.00464$ | This Work |
| $T_{eff}$ (K) | $3390 \pm 44$ | This Work |
| **TOI 700 b** | | |
| $T_0(BJD - 2457000)$ | $1490.98 \pm 1$ | Exo-FOP |
| $\ln(Period\ [days])$ | $2.300 \pm 1$ | Exo-FOP |
| $\ln(Transit\ Depth\ [ppm])$ | $6.221 \pm 0.095$ | Exo-FOP |
| **TOI 700 c** | | |
| $T_0(BJD - 2457000)$ | $1548.75 \pm 1$ | Exo-FOP |
| $\ln(Period\ [days])$ | $2.776 \pm 1$ | Exo-FOP |
| $\ln(Transit\ Depth\ [ppm])$ | $8.099 \pm 0.073$ | Exo-FOP |
| **TOI 700 d** | | |
| $T_0 (BJD - 2457000)$ | $1742.15 \pm 1$ | Exo-FOP |
| $\ln(Period\ [days])$ | $3.622 \pm 1$ | Exo-FOP |
| $\ln(Transit\ Depth\ [ppm])$ | $6.418 \pm 0.073$ | Exo-FOP |

*2.4: Lightcurve fitting and analysis*

The reduced and extracted lightcurves are fit for transits using *exoplanet* and its dependencies (Foreman-Mackey et al. 2021; Salvatier et al. 2016; Theano Development Team et al. 2016; Kumar et al. 2019; Astropy Collaboration et al. 2018; Luger et al. 2019; Foreman-Mackey et al. 2017). The stellar mass and radius are parameterized via a normal distribution bounded above 0. Limb-darkening parameters are estimated in-model following Kipping (2013a) and were sampled uniformly. The impact parameters were sampled uniformly between zero (the center of the star) and one plus the planet-to-star radius ratio for each planet (the edge of transit overlap). We used a beta distribution as a prior on the eccentricities as suggested by Kipping (2013b), bounded between zero and one. The periastron angle was sampled randomly. Transit depths and periods were parameterized using their natural logs. All other system parameters are modeled using a Gaussian distribution.

The list of parameters included in the fit as well as the values is available in Table 1. Parameters described above that are not included in the table had random initial values. Variance for some parameters in the prior distributions is intentionally left high so as to not unduly influence the fitting procedure.

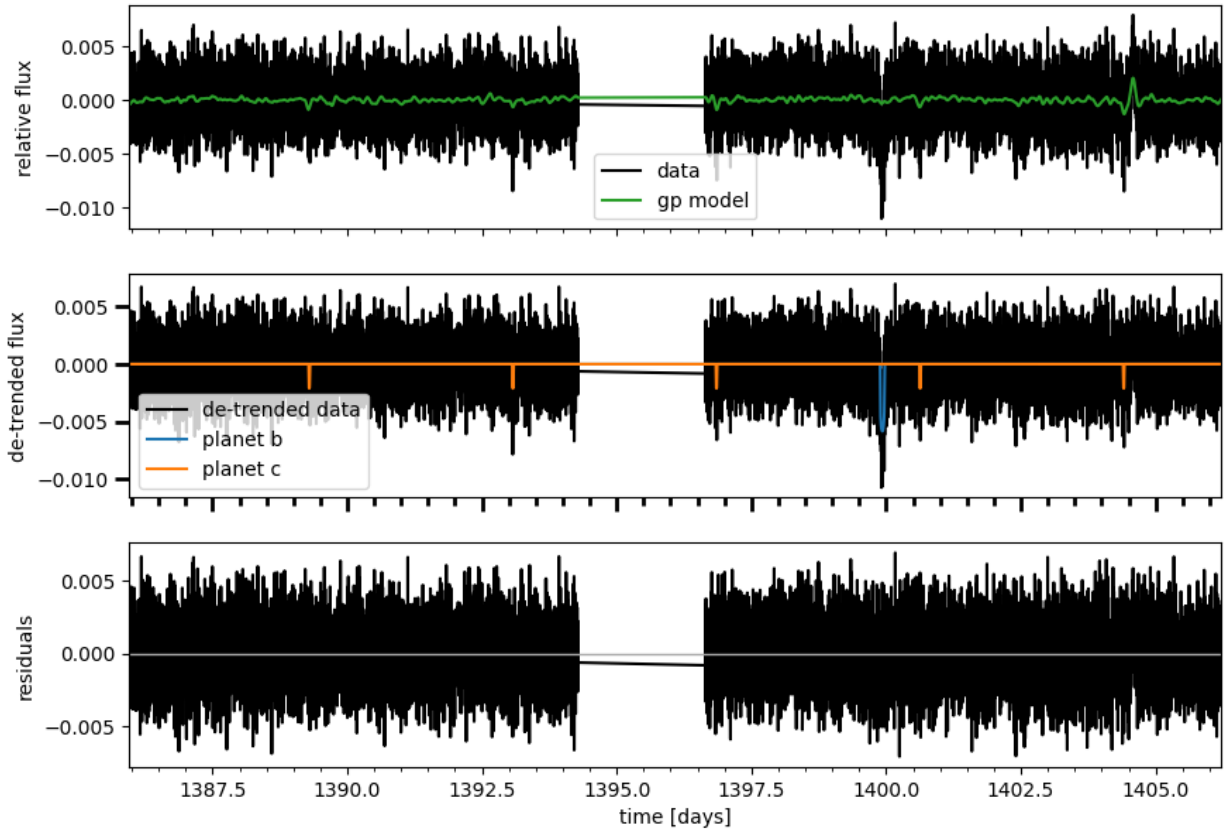

**Figure 2.** *TESS* lightcurve for TOI 256 cleaned for instrument systematics and normalized (upper panel) showing the Gaussian Process (GP) model (green curve), lightcurve minus the GP model (middle panel) showing the *exoplanet* transit model for planets b (blue curve) and c (orange curve), and the residual data with the GP and transit models subtracted (bottom panel).

To model noise, we fit a Gaussian process (GP) that is a mixture of two simple harmonic oscillators comprised of two hyperparameters, $ln(\rho_{GP})$ and $ln(\sigma_{GP})$, as well as a mean flux offset, $\mu_F$, and the natural log of the variance, $ln(\sigma_F)$. The initial value for $\mu_F$ is 0 and $\sigma_F$ is calculated from the variance of the flux with transits masked. The initial value for $ln(\sigma_{GP})$ is set to be $ln(\sigma_F)$ and $ln(\rho_{GP})$ is initially 0. Variance for the noise parameters is set at 10 for each value to allow adequate flexibility in the modeling process.

We first run the parameters through the version of the *PyMC3* (Salvatier et al. 2016) optimization routine which was made for use with *exoplanet*. The initial transit lightcurves for each planet are plotted along with the Gaussian process model and residual (Figure 2) and are visually inspected to confirm the model transits line up with the observed lightcurve. In addition, the optimization solution is inspected for significant deviations from the expected parameters as this can indicate that the model is incorrect and must be changed. In practice, changes are made if the median values differ from expected by more than a standard deviation, the errors are abnormally large, or the model chains do not converge.

After the initial model is inspected, a mask is created for data outside 5 $\sigma$ of the residual mean. This secondary data clipping enables more accuracy in the Gaussian process modeling, resulting in a more accurate lightcurve fit.

The resulting lightcurve is used in a second optimization of the parameters which yields a final model to be sampled using the No U-turn Sampler (Hoffman & Gelman 2011). We use 4 chains with 6000 tuning steps and 5000 draw steps each, yielding a trace of 20000 steps total. For all traces, the Gelman-Rubin diagnostic (Gelman & Rubin 1992) was near 1.0 and the number of effective samples was over 1000 for all parameters. The sampled traces are saved as fits files and are available upon request. Section 4 shows the deterministic parameters computed as part of this sampling as well as their median and variance.

## 3. DESICCATION SIMULATIONS

A major concern of exoplanet habitability is the ability to sustain liquid water at the surface long enough for life to have the chance to form, especially around smaller (i.e. M-dwarf) stars (Luger & Barnes 2015). These smaller stars tend to actively flare well into their lives, spewing dangerous amounts of XUV radiation that can destroy water and lead to the total desiccation of their satellites (Reid & Hawley 2005; Scalo et al. 2007).

On Earth, our one data point for life, the period between the formation of the oceans and the initial stage of life is estimated to be at most 770 Myr and as little as 260 Myr (Dodd et. al 2017). Because of the activity level of M-dwarf stars, however, we will investigate whether or not the hydrosphere is stable up to 1 Gyr after the system forms. Though we cannot say for certain if life has or will evolve on the targets, this gives an upper estimate on the ability of the targets to sustain the necessary environment for life as we know it.

For each parameter set in the trace obtained by the methods in Section 2, we instantiate and execute two *VPlanet* simulations of orbital, rotational, and atmospheric evolution using the *stellar*, *atmesc*, and *eqtide* modules. The first of the two simulations assumes the planet formed roughly 100 Myr after the star and integrates until 1Gyr. The second simulation takes the system environment as it has been recently observed and models the next Gyr. Both simulations address the issue of hydrosphere stability, focusing on primary and secondary atmospheres, respectively.

*3.1: VPlanet setup and assumptions*

Before setting up the input files, we estimate the planet mass using the non-parametric method from Ning et al. (2018). We use the *mrexo* package (Kanodia et al. 2019) for *python* with the option to generate a complimentary MCMC trace for a posterior sample of the target planets' radii (Figure 3). A practical example can be seen in (Figure 4) for TOI 256 b.

As fluctuation in the stellar luminosity can greatly affect the model output, we chose to use the accepted values of 0.0233 (solar luminosity) for TOI 700 and 0.00441 (solar luminosity) for TOI 256 instead. Current stellar age is also taken to be the accepted values of 1.5 Gyr and 5 Gyr respectively. The radius and mass (and thus the density) are fixed to the average values of the trace. As is standard in these simulations for M-dwarf type stars, we use 1e-3 F_XUV/F_BOL for the saturation phase and $\tau = 1$ Gyr (Ribas et al. 2005). All other assumptions made in the VPlanet model are held excepting the variables mentioned.

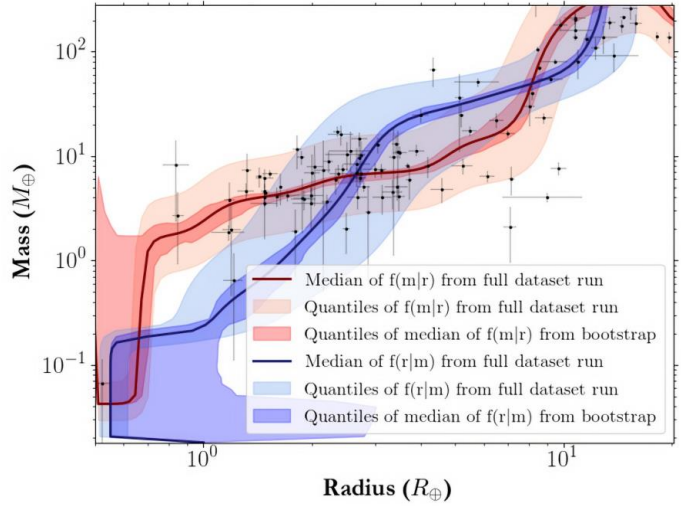

**Figure 3.** Figure adapted from Kanodia et al. (2019) highlighting the nonparametric fit for mass given radius used in this work shown in red. The accompanying fit for radius given mass is also shown in blue. The mean of the fits is shown as a dark line and the shaded regions show the 16% and 84% quantiles for the conditional distributions. Data used in this fitting are from the *Kepler* dataset.

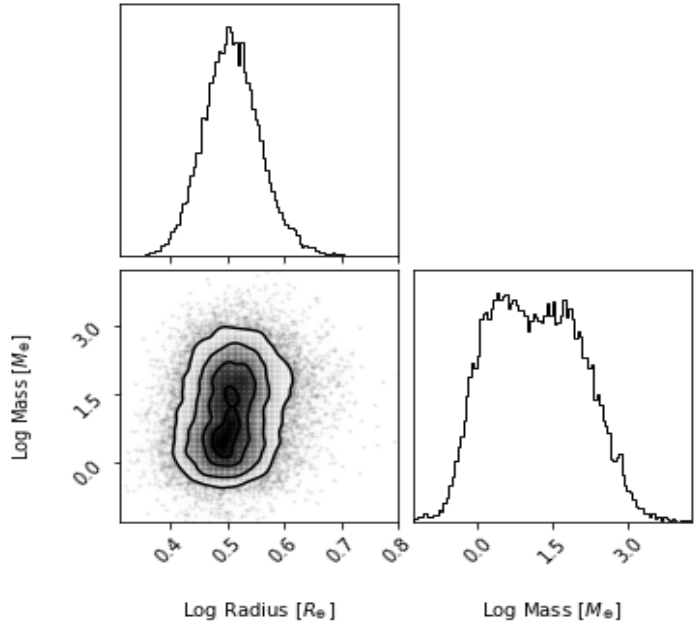

**Figure 4.** Log radius (upper left panel) from the *TESS* data fitting, log mass given this radius distribution (bottom right panel), and their 2D probability distribution (bottom left panel) showing the bimodality of the mass fitting.

In order to simulate the hydrosphere degradation during the active phase of the star, we assign the initial water on the simulated planet to be 10 Earth Oceans as in Barnes et al. (2018) for their most water-rich case. As we lack planetary formation and migration information of the target systems, their initial water inventory is difficult to ascertain. However, Earth is thought to have formed and accreted up to 70 oceans worth of water (Morbidelli et al. 2000; Raymond et al. 2006; Chassefière et al. 2012), making this a conservative estimate for initial water inventory.

The stability of a secondary hydrosphere is tested by setting the initial water inventory in Earth oceans to be proportional to the planet's mass in Earth masses. This tests the possibility that the planets can currently support a proportional amount of stable water for a long time and thus retain any primary or secondary hydrosphere for long enough that life could develop. Neither method is intended to be a definitive judgement of the planets' habitability.

The simulation results are given in Section 4. For statistics relating to TOI 700 d, the parameters are weighted according to the number of data points in their respective lightcurves. The output files and aggregate files of the final parameters are available upon request.

*3.2: Desiccation metrics*

We use four desiccation metrics to analyze the VPlanet outputs: desiccation percentage, remaining water inventory, percent of static models, and non-static desiccation rate. The necessary components of each metric are calculated during the data aggregation phase along with the results presented in Section 4.

*3.2.1: Desiccation Percentage*

This metric is simply calculated by dividing the number of simulations that become completely desiccated by the total number of simulations. As the uncertainty relating to the fitted parameters could be relatively large, this metric gives a simple first glance at the sustainability of the hydrosphere across the parameter space. Desiccation percentage is mostly applicable to primary hydrosphere degradation as this is when the bulk of water is lost in the simulations.

*3.2.2: Remaining Water Inventory*

The remaining water inventory is calculated from the water mass of the final time step in the simulation. For the set of simulations starting with 10 Earth oceans of water, the metric is taken to be this final value. This is instead expressed as a percentage of the original water mass for the secondary simulations with proportional initial water. Two versions of this metric are presented here: one including desiccated cases and one that does not. Including desiccated cases essentially yields a reflection of the desiccation percentage metric, while rejecting desiccated cases explores not only the final water distribution in hydrated cases, but, more importantly, what the expected amount of water is given water is discovered on the target.

*3.2.3: Static Cases*

The static case percentage tracks how many of the final model states have ceased losing water. As desiccated models would inherently be static, they are excluded. To calculate the metric, the difference of water present in the final two steps of non-desiccated models is taken and counted if it is zero. After the aggregation is finished, the final count of static cases is divided by the number of non-desiccated models to form the metric. This reveals the stability of the planets' overall hydrosphere, should one exist, which is important for determining longer-term sustainability.

*3.2.4: Non-Static Desiccation Rate*

Finally, for those models that are losing water, we take a derivative 1Myr from the stop time to determine the final desiccation rate. This is a fairly selective metric as in both systems the majority of cases were static, however, in more active systems, this metric is important to determine the likelihood of complete desiccation at the present time.

## 4. RESULTS

*4.1: System Parameters*

In this section, we present the deterministic parameters computed by sampling the resultant trace of fitting TESS lightcurves for TOIs 700 and 256 as well as their median and variance. In addition, the naive mass prediction for each planet is given.

TOI 700 d and 256 b have posterior predicted periods of $P_{700\,d} = 37.406\,^{+0.067}_{-0.061}$; $P_{256\,b} = 24.737\,^{+6.33e-5}_{-6.12e-5}$ days and radius of $R_{700\,d} = 1.123\,^{+0.131}_{-0.114}\,R_\oplus$; $R_{256\,b} = 1.650\,^{+0.087}_{-0.078}\,R_\oplus$. The results of the fit are well in-line with other studies using *TESS* data (Gilbert et al. 2020; Lillo-Box et al. 2020; Exo-FOP 2019). Gilbert et al. (2020) and Lillo-Box et al. (2020) use other observations to model stellar variability from the All-Sky Automated Survey for Supernovae and HARPS/ESPRESSO respectively, leading to small discrepancies between their studies and this work.

From radial velocity observations, the mass of TOI 256 b is known to be $M_{256\,b} = 6.48\,\pm\,0.46\,M_\oplus$ (Lillo-box et al. 2020), well within the standard error on our naïve prediction, presented in Table 2, which also considers the potential for the target to be gaseous. The effect of the bimodal mass predictions is evident in Figures 4.1, 4.2, and 4.3 where the predicted density has a degeneracy with predicted water loss. The values given here and used in the simulations serve as examples for when radial velocity measurements are not available.

*4.2: Desiccation Metrics*

We present the desiccation metrics for TOIs 700 d and 256 b in Table 3 as well as the correlation between water loss and density in Figures 5 and 6. In many cases, both targets appear to be able to maintain a stable hydrosphere which is a strong predictor for their habitability; all cases with terrestrial densities retain some water.

For all planets in the target systems, modern simulations using proportional water inventories resulted in no water loss. This is due in majority to the low activity levels of the host stars, resulting in low XUV flux. Thus, these results are not presented in Table 3. Unfortunately, no other planet besides the two targets shown in Table 3 had more than 1% of their simulations conclude with remaining water. In addition, of those that retained water, none had any static cases, meaning their hydrospheres would be predicted to degrade until fully desiccated. They are excluded from the table for this reason.

Desiccated models are included when calculating remaining water inventory statistics, resulting in a peak in the distribution at zero. Because of this, the desiccation percent is a more useful metric for assessing the stability of a primary/formation hydrosphere. As expected, the planets that are safely in the habitable zone have the fewest desiccated cases and retain the most water.

*4.2.1: TOI 700d*

As just over 50% of models for TOI 700 d returned a desiccated state, the median value of remaining water mass is 0 Earth oceans. For this reason, we also present this metric without desiccated values to examine the distribution of remaining water mass should the planet retain water. This resulted in $4.69^{+2.53}_{-2.72}$ Earth oceans remaining. For TOI 700d and smaller planets, this non-zero value is more appropriate to consider as the prospective mass of desiccated cases is inconsistent with reasonable density for a rocky planet.

Table 2. Planet Parameters

| Parameter | Median | $-1\sigma$ | $+1\sigma$ |
|---|---|---|---|
| *TOI 700* | | | |
| **TOI 700 b** | | | |
| Period (days) | 9.977 | 0.003 | 0.019 |
| Radius ($R_\oplus$) | 0.980 | 0.090 | 0.091 |
| $T_0$ (BJD – 2457000) | 1331.263 | 0.652 | 0.075 |
| Impact Parameter | 0.242 | 0.176 | 0.262 |
| ln($depth$) | 6.276 | 0.140 | 0.136 |
| Mass ($M_\oplus$) | 0.840 | 0.577 | 1.257 |
| Mass (from log) | 1.206 | 0.733 | 1.199 |
| **TOI 700 c** | | | |
| Period (days) | 16.045 | 0.006 | 0.005 |
| Radius ($R_\oplus$) | 2.975 | 0.185 | 0.271 |
| $T_0$ (BJD – 2457000) | 1339.991 | 0.084 | 0.072 |
| Impact Parameter | 0.908 | 0.020 | 0.018 |
| ln($depth$) | 8.061 | 0.047 | 0.050 |
| Mass ($M_\oplus$) | 5.377 | 4.392 | 12.072 |
| Mass (from log) | 7.480 | 4.043 | 11.273 |
| **TOI 700 d** | | | |
| Period (days) | 37.406 | 0.062 | 0.068 |
| Radius ($R_\oplus$) | 1.123 | 0.114 | 0.131 |
| $T_0$ (BJD – 2457000) | 1330.401 | 0.560 | 0.485 |
| Impact Parameter | 0.424 | 0.278 | 0.233 |
| ln($depth$) | 6.501 | 0.166 | 0.167 |
| Mass ($M_\oplus$) | 0.922 | 0.660 | 1.575 |
| Mass (from log) | 1.404 | 0.681 | 1.492 |
| **TOI 256 b** | | | |
| Period (days) | 24.737 | 6.1e-5 | 6.3e-5 |
| Radius ($R_\oplus$) | 1.650 | 0.078 | 0.087 |
| $T_0$ (BJD – 2457000) | 1399.929 | 0.001 | 0.001 |

| Parameter | Median | $-1\sigma$ | $+1\sigma$ |
|---|---|---|---|
| Impact Parameter | 0.269 | 0.188 | 0.232 |
| ln($depth$) | 8.806 | 0.053 | 0.052 |
| Mass ($M_\oplus$) | 2.320 | 1.822 | 5.221 |
| Mass (from log) | 2.847 | 1.746 | 4.928 |
| **TOI 256 c** | | | |
| Period (days) | 3.778 | 6e-6 | 6.1e-6 |
| Radius ($R_\oplus$) | 1.155 | 0.057 | 0.063 |
| $T_0$ (BJD – 2457000) | 1389.293 | 0.001 | 0.001 |
| Impact Parameter | 0.237 | 0.173 | 0.267 |
| ln($depth$) | 8.103 | 0.055 | 0.054 |
| Mass ($M_\oplus$) | 1.103 | 0.700 | 1.445 |
| Mass (from log) | 1.413 | 0.636 | 1.371 |

*4.2.1: TOI 256 b*

All simulations of TOI 256 b (LHS 1140 b) returned non-desiccated models with $6.73^{+1.14}_{-0.62}$ Earth oceans of water remaining. As this planet has a mass consistent with a rocky planet (Dittman et al. 2017; Lillo-Box, 2020), these simulations indicate that LHS 1140 b could likely have a large ocean covering most if not all of the surface as has been previously predicted (e.g., Lillo-Box et al. 2020).

*4.3: Discussion*

*4.3.1: Mass-dependance for Rocky Bodies*

Figure 5 illustrates the dependance of water retention on the mass of the planet for TOI 700 d. At masses lower than the median predicted mass in log space, ∼1.4 $M_\oplus$, the planet is certainly desiccated, but at masses indicating a terrestrial density, the planet only loses a maximum of 8 Earth oceans of water. As expected, more massive planets retain more water as their surface gravity is higher. The example of TOI 700 d clearly shows the necessity of thorough radial velocity measurements of smaller exoplanets as density, and thus mass, are such strong predictors of hydrosphere stability.

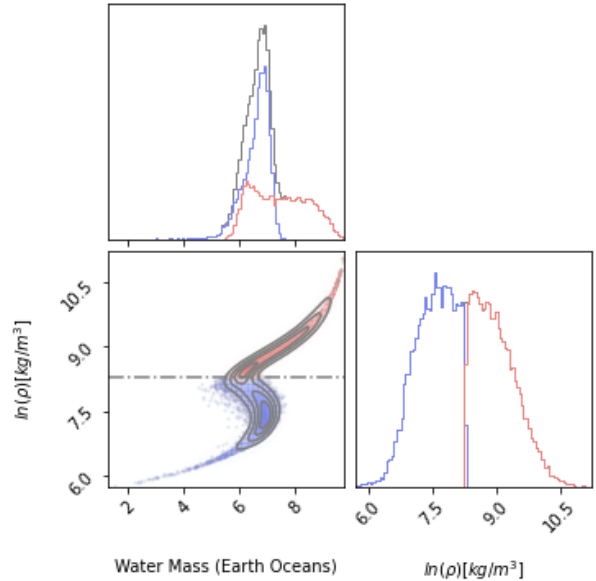

**Figure 5.** Remaining water mass (upper left panel), natural log of the predicted density (bottom right panel), and their 2D probability distribution (bottom left) for TOI 256 b. The data is split into higher density (pink) and lower density (blue) for all panels using the density of Mars (horizontal line in bottom left panel) as the dividing factor. The increase in water retention below the density division is due to the implied presence of an extended gaseous envelope. As the known natural log density of TOI 256 b is approximately 8.96 (Lillo-Box et al. 2020), we predict that it will retain most of its initial water inventory.

**Table 3.** Desiccation Metrics

| Metric | Median | −1σ | +1σ | Metric | Median | −1σ | +1σ |
|---|---|---|---|---|---|---|---|
| | TOI 700 | | | | TOI 256 | | |
| Desiccation % | 0 % | … | … | Desiccation % | 0.508 | … | … |
| Remaining Water Inventory (Earth Oceans) | 6.733 | 0.622 | 1.138 | Remaining Water Inventory (Earth Oceans) | 0 | 0 | 5.811 |
| Non-desiccated Inventory (Earth Oceans) | … | … | … | Non-desiccated Inventory (Earth Oceans) | 4.694 | 2.719 | 2.531 |
| Final Static % | 100 % | … | … | Final Static % | 0.613 | … | … |
| Desiccation Rate (Earth Oceans/Gyr) | … | … | … | Desiccation Rate (Earth Oceans/Gyr) | 2.602 | 0.805 | 1.600 |

*4.3.2: Density Dependance for Intermediate Bodies*

Figure 6 presents the desiccation of LHS 1140 b against the predicted density used in the model. This is similar to the previous figure, but as this planet's radius is in the Fulton Gap (Fulton et al. 2017), a naïve mass approximation gives a relatively equal probability of it being rocky or gaseous. The density of Mars is used as a dividing line, highlighting the differences between these types of planets.

For rocky planets, the curve follows much the same curve as TOI 700 d. As the density and thus the mass lowers, the planet is unable to retain the photodisassociated water and becomes desiccated. However, for gaseous planets, the amount of water retained increases with decreasing density for a portion of the distribution. This is due to the implied presence of an extended gaseous envelope that both protects the lower atmospheric water from disassociation and prevents Hydrogen and Oxygen from outflowing effectively. Similarly to the previous target, the case of TOI 256 b highlights the usefulness of radial velocity measurements in determining the potential habitability of an exoplanet.

*4.3.3: Comparison of TOI 700 d to Other Studies*

Suissa et al. (2020) explored 20 possible climate scenarios for TOI 700 d using the ExoCAM modeling package. Of these scenarios, the most likely ones given the work in our study are aqua planets in synchronous orbits. As we predict the loss of a significant amount of water, this correlates with the buildup of a relatively large amount of O2 on the planet's surface (Luger & Barnes 2015). As these simulations do not include a photochemical model, they neglect this significant factor. Because of this, more simulations including photochemical modeling must be conducted to predict the potential atmospheric states more accurately.

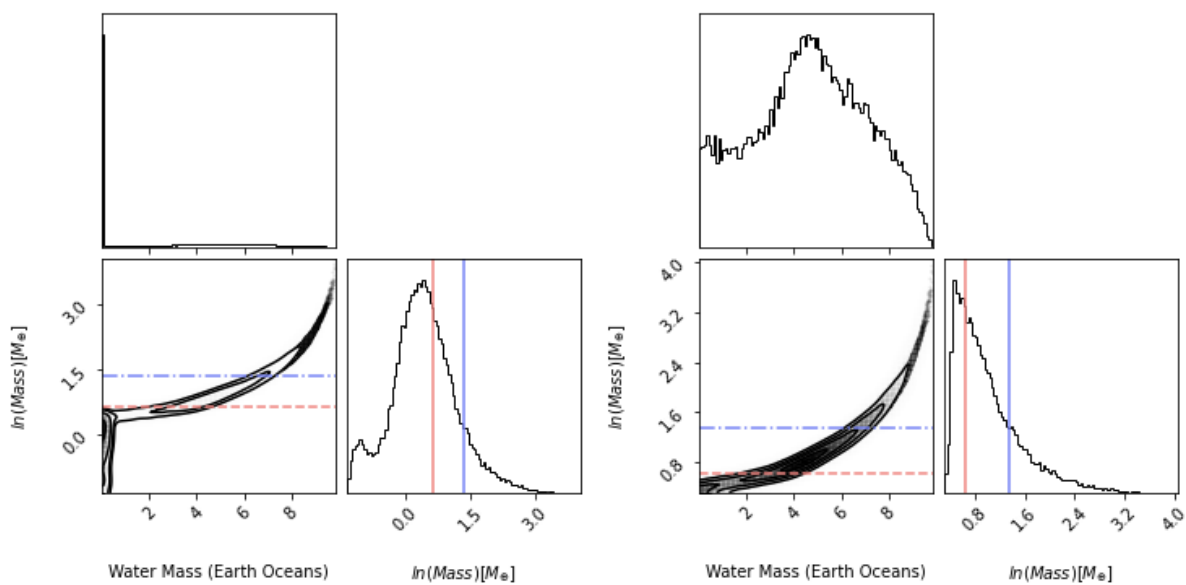

**Figure 6.** The left figure shows the remaining water inventory (upper left panel), natural log of the predicted mass (bottom right panel, and their 2D probability distribution (bottom left) for all predicted masses. The right figure shows the same parameters, but only for the cases where the model did not become fully desiccated. On both figures, the natural log of the mass of Mars is shown as a pink line and the mass predicted by the equations in Sotin et al. (2007) is shown as a blue line.

*4.3.4: Future Observations*

Unfortunately, TOI 700 will not be observed by JWST in its Cycle 1 observations, potentially due to its small size and thus signal-to-noise ratio. Hopefully, the potential existence of a fourth planet in the system will lead to TOI 700 becoming a higher priority system. Radial velocity measurements of TOI 700 d would greatly aid in further characterization of the planet and give a foundation for the planning of atmospheric observations with JWST.

Water on LHS 1140 b was tentatively discovered using HST's WFC3 (Edwards et al., 2020). This, along with radial velocity observations using ESPRESSO (Lillo-Box et al., 2020) strongly implies the existence of a large surface ocean on the planet which is consistent with our findings that LHS 1140 b should lose little, if any, of its primordial and accumulated water. Observations of LHS 1140 b in Cycle 1 of JWST are aimed at confidently identifying water and other volatiles present in the atmosphere of the planet.

## 5. CONCLUSIONS

We have analyzed two high priority *TESS* planets, TOIs 256 b and 700 d, for their potential to retain a stable hydrosphere using a new, flexible method that incorporates observational data as opposed to other methods which use assumed parameter distributions. Preserving the correlations inherent in data-fitting procedures, such as between the fitted radius and mass predictions, is an important step in assessing the habitability of exoplanets and deciding which, if any, future observations are needed for further refinement.

We applied the method to two highly prioritized *TESS*-observed planets, TOIs 256 b (LHS 1140 b) and 700 d and assessed the likelihood that they retain stable hydrospheres to the present day through several metrics. Even with conservative initial water inventories, both planets retain water in a significant number of modeled cases (100% and 49% respectively). In addition, both planets are highly likely to retain a significant amount of water if they are predominately rocky.

Further observations of both planets are upcoming priorities, with LHS 1140 b to be observed by the recently launched *JWST* in its first Cycle. The characterization of this planet's atmosphere will enable more precise simulations of other planets in the future, refining procedures for identifying and prioritizing potentially habitable planets.

Though TOI 700 d is not a target for *JWST's* Cycle 1, it remains a tantalizing target for radial velocity observations as well as more precise photochemical modeling. The difficulty of atmospheric observations of this planet is a concern. However, the potential for discovering a habitable Earthlike planet is undeniable.

*This material is based upon work supported by the National Aeronautics and Space Administration under a FINESST Award No. 80NSSC20K1395 issued through the Mission Directorate.*

*This research is supported by the Arkansas High Performance Computing Center, multiple National Science Foundation grants, and the Arkansas Economic Development Commission.*

*This paper includes data collected by the TESS mission. Funding for the TESS mission is provided by the NASA's Science Mission Directorate.*

*This research has made use of the Exoplanet Follow-up Observation Program website, which is operated by the California Institute of Technology, under contract with the National Aeronautics and Space Administration under the Exoplanet Exploration Program.*

*This publication makes use of data products from the Two Micron All Sky Survey, which is a joint project of the University of Massachusetts and the Infrared Processing and Analysis Center/California Institute of Technology, funded by the National Aeronautics and Space Administration and the National Science Foundation.*

*This work has made use of data from the European Space Agency (ESA) mission Gaia (https://www.cosmos.esa.int/gaia), processed by the Gaia Data Processing and Analysis Consortium (DPAC, https://www.cosmos.esa.int/web/gaia/dpac/consortium). Funding for the DPAC has been provided by national institutions, in particular the institutions participating in the Gaia Multilateral Agreement.*